# Quantum Teleportation within a Quantum Network

**Hari Prakash[1,2], Ajay K Maurya[1] and Manoj K Mishra[1]**

[1]Physics Department, University of Allahabad, Allahabad, India
[2]Indian Institute of Information Technology, Allahabad, India
Emails: *prakash_hari123@rediffmail.com, ajaymaurya.2010@gmail.com*

and *manoj.qit@gmail.com*

**Abstract**: We consider the problem of teleporting an unknown information state within a quantum network by a sender, *say*, Alice to any given receiver out of several receivers, *say*, Bob[(1)], Bob[(2)],…, Bob[(n)]. For this task, we suggest two schemes using ($n$+1)-partite GHZ state as a quantum channel. The first scheme involves C-NOT operations on Bob's qubit with the control as Alice's qubit, while the second scheme involves measurements in the Hadamard basis by Bobs on their qubits. We have generalized both schemes for quantum teleportation of arbitrary $m$-qubit information state within the quantum network. Further, we have proposed two experimental schemes for the generation of ($n$+1)-partite GHZ state using interaction between highly detuned Λ-type three-level atoms and optical coherent field. We also discuss the possible experimental imperfections like, atomic-radiative time, cavity damping time and atom-cavity interaction time.

## 1 Introduction

Quantum teleportation (QT), first due to Bennett *et al* [1], involves complete transfer of an unknown quantum state from one quantum mechanical particle to another particle across space using long range EPR correlations (quantum entanglement) [2] and a classical channel. Since then, a number of theoretical [3-4] and experimental [5-9] studies on QT have been done. Several authors have experimentally demonstrated QT of unknown single photon polarized state using standard bi-photonic Bell states [5-6], atomic qubit using bipartite atomic Bell states [7-8] and unknown quantum state of nucleus [9].

Further, the idea of QT of single qubit information state has been extended by introducing tripartite entangled states as the quantum channel and a third party Charlie between the two legitimate users, Alice and Bob. This third party Charlie controls the whole process of QT and hence this QT is called controlled QT (CQT). Some authors proposed schemes for CQT of single qubit information state using GHZ state [10] or GHZ class states [11], while authors [12-13] proposed schemes for controlled QT of single qubit information state using W state.

Rigolin [14] defined a set of 16 mutually orthogonal states of four qubits and used them for faithful QT of two qubit information state. However, Deng [15] has pointed that Rigolin's states are tensor product of two standard bipartite Bell states and hence this protocol is principally equivalent to the Yang-Guo protocol [16]. In practical situations, there may be need of sending large amount of information encoded in multi-qubits. Hence the QT protocol has been generalized further for multi-qubit information state by a number of authors [17-19]. Chen *et al* [17] proposed QT of an arbitrary *n*-qubit information state using the generalized 2*n*-qubit Bell states, which are simply tensor product of *n* standard bipartite Bell states.

To increase the security of multi-qubit QT schemes, some authors [18-19] proposed the CQT of unknown multi-qubit information states by introducing a third observer or a number of observers between two legitimate observers, Alice and Bob. These schemes of QT allow the transfer of information from Alice to Bob, *i.e.*, one way quantum communication. The authors [20] proposed idea of two-way quantum communication termed as secure quantum information exchange that enables the secure exchange of information states between Alice and Bob simultaneously with the classical assistance of a third party Charlie.

The goal of quantum information science is to realize the quantum networks that are composed of many nodes and channels. It requires the quantum systems that are ideal for long-term storage of the quantum information and also the quantum channels that allows the fast transfer of quantum information between the nodes. It is expected that QT can be effectively used as the link between two quantum processors (nodes) working distant apart in a quantum network. So it is required to have a scheme that can set a quantum communication link between any two legitimate users of the quantum network.

In this context, we consider a quantum network composed of (*n*+1) observers, a sender, say, Alice and *n* receivers, say, Bob$^{(1)}$, Bob$^{(2)}$,...., Bob$^{(n)}$. Our aim is to teleport single qubit information state from Alice to anyone out of *n* receivers (Bobs). For this purpose, we suggest two schemes using (*n*+1)-partite GHZ state as quantum channel shared among the (*n*+1) observers. The first scheme involves C-NOT operations on Bob's qubit with control as Alice's qubit and the second scheme involves a measurement in the basis $|A_{\pm}\rangle = \frac{1}{\sqrt{2}}[|0\rangle \pm |1\rangle]$ by Bobs on their qubits. We have generalized both schemes for QT of multi-qubit information state within the quantum network.

For the experimental realization, we propose two efficient schemes for the generation of (*n*+1)-qubit GHZ state using interaction between highly detuned Λ-type three-level atoms and optical coherent field. Further, we discuss about various experimental parameters, *e.g.*, atomic-radiative time, cavity damping time and atom-cavity interaction time and show that our schemes can be realized with the presently available technology.

**2 Quantum Teleportation of single qubit information state within a Quantum Network**

Consider a quantum network consisting of (*n*+1) observers, *say,* Alice and *n* numbers of Bobs, sharing an (*n*+1)-partite GHZ state of the form,

$$|E\rangle_{A,B1,B2,....,Bn} = \frac{1}{\sqrt{2}}[|000....0\rangle + |111....1\rangle]_{A,B1,B2,....,Bn}, \qquad (1)$$

where $|0\rangle$ and $|1\rangle$ are ground and excited levels of a two-level system, subscripts represent the quantum modes. The entangled mode 'A' belongs to Alice (the sender) while the entangled modes $B1, B2, ..., Bn$ are sent to Bob$^{(1)}$, Bob$^{(2)}$, ...., Bob$^{(n)}$ (the receivers) respectively. Let us consider that Alice wishes to teleport single qubit information state in mode 'I' given by

$$|I\rangle_I = [a_0|0\rangle + a_1|1\rangle]_I , \qquad (2)$$

to any one among the $n$ Bobs. Alice mixes her entangled particle in mode 'A' with information mode 'I'. Initial state of the system can be written as

$$|\psi\rangle_{I,A,B1,B2,....,Bn} = |I\rangle_I \otimes |E\rangle_{A,B1,B2,....,Bn}. \qquad (3)$$

Let Alice decides to teleport the information state (2) to the $i^{th}$ Bob, where $i$ belongs to (1, 2, ..., $n$). For completion of this task, we propose two schemes.

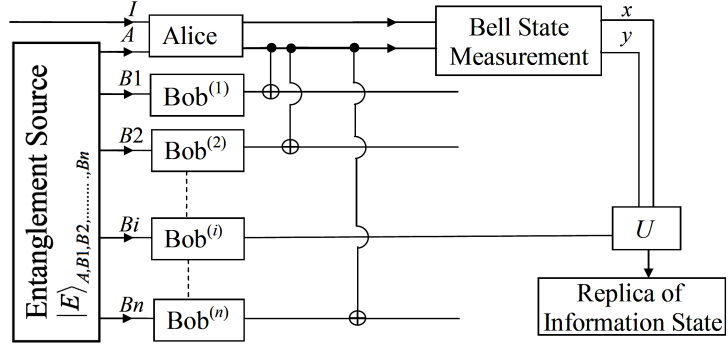

**Figure 1:** Scheme for QT of single qubit information state to $i^{th}$ Bob within a quantum network (First scheme). Information mode $I$ and Entangled mode $A$ belong to Alice, while entangled mode $Bi$ ($i=1, 2, ...,n$) is with $i^{th}$ Bob. $x$ & $y$ stand for classical bits. $U$ refers to the unitary operation to be performed by $i^{th}$ Bob for completing the QT.

In the first scheme (see Figure **1**), Alice asks all Bobs except the $i^{th}$ Bob to perform C-NOT operation on their particles with control as her qubit in mode 'A'. After this process the state (3) can be written as

$$|\psi\rangle_{I,A,B1,B2,....,Bn} = |I\rangle_I \otimes \frac{1}{\sqrt{2}}[|00\rangle + |11\rangle]_{A,Bi} \otimes |00....0\rangle_{B1,B2,...,B(i-1),B(i+1),....,Bn} . \qquad (4)$$

From equation (4), it is clear that all Bobs, except the $i^{th}$ Bob, get disentangled from Alice. The state (4) can also be written as

$$|\psi\rangle_{I,A,B1,B2,....,Bn}$$

$$= \frac{1}{2}[|B\rangle_{I,A}^{(0)} \otimes (a_0|0\rangle + a_1|1\rangle)_{Bi} + |B\rangle_{I,A}^{(1)} \otimes (a_0|0\rangle - a_1|1\rangle)_{Bi} + |B\rangle_{I,A}^{(2)} \otimes (a_1|0\rangle + a_0|1\rangle)_{Bi}$$

$$+ |B\rangle_{I,A}^{(3)} \otimes (a_1|0\rangle - a_0|1\rangle)_{Bi}] \otimes |00....0\rangle_{B1,B2,...,B(i-1),B(i+1),....,Bn}, \quad (5)$$

where

$$|B\rangle_{I,A}^{(0,1)} = \frac{1}{\sqrt{2}}[|00\rangle \pm |11\rangle]_{I,A} \text{ and } |B\rangle_{I,A}^{(2,3)} = \frac{1}{\sqrt{2}}[|01\rangle \pm |10\rangle]_{I,A},$$

are the standard bipartite Bell states. Now Alice performs Bell state measurement (BSM) in the standard bipartite Bell basis and conveys the BSM result to $i^{th}$ Bob through a classical 2-bit channel. On the basis of measurement result conveyed, $i^{th}$ Bob performs suitable unitary transformation to get faithful replica of the original information state. From equation (5), it is clear that $I$, $\sigma_z$, $\sigma_x$ and $\sigma_z \sigma_x$ are the unitary transformations performed by $i^{th}$ Bob for the result 0, 1, 2 and 3 of the Alice's BSM respectively.

In the second scheme (see Figure **2**), Alice asks to all Bobs except the $i^{th}$ Bob to perform projective measurement on their particles in the basis $|A_\pm\rangle = \frac{1}{\sqrt{2}}[|0\rangle \pm |1\rangle]$ and convey their results back. Let the result of measurement of $r$ number of Bobs is $|A_-\rangle$ and rest ($n$-$r$-1) Bob is $|A_+\rangle$. Then the state of the whole system (3) becomes,

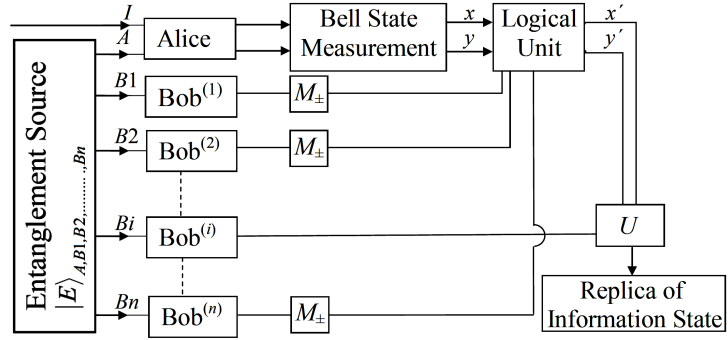

**Figure 2:** Scheme for teleportation of single qubit information to $i^{th}$ Bob within a network (Second scheme). Information mode $I$ and Entangled mode $A$ belong to Alice, while entangled mode $Bi$ ($i$=1, 2, …,$n$) is with $i^{th}$ Bob. $x$, $y$, $x'$ & $y'$ stand for classical bits. $M_\pm$ refer to measurement in basis $|A_\pm\rangle = \frac{1}{\sqrt{2}}[|0\rangle \pm |1\rangle]$. $U$ refers to the unitary operation to be performed by $i^{th}$ Bob for completing the QT.

$$|\psi\rangle_{I,A,B1,B2,....,Bn} = |I\rangle_I \otimes \frac{1}{\sqrt{2}}[|00\rangle + (-1)^r|11\rangle]_{A,Bi} \qquad (6)$$
$$\otimes [\prod_r |A_-\rangle \otimes \prod_{n-r-1}|A_+\rangle]_{B1,B2,...,B(i-1),B(i+1),...,Bn},$$

where $\prod_r$ indicates direct product of $r$ identical factors.

From equation (6), it is clear that all Bobs, except the $i^{th}$ Bob, get disentangled from Alice. Using the standard bipartite Bell basis, the combined state (6) can be rewritten as

$$|\psi\rangle_{I,A,B1,B2,....,Bn}$$
$$= \frac{1}{2}[|B\rangle_{I,A}^{(0)} \otimes (a_0|0\rangle + (-1)^r a_1|1\rangle)_{Bi} + |B\rangle_{I,A}^{(1)} \otimes (a_0|0\rangle - (-1)^r a_1|1\rangle)_{Bi} \qquad (7)$$
$$+ |B\rangle_{I,A}^{(2)} \otimes (a_1|0\rangle + (-1)^r a_0|1\rangle)_{Bi} + |B\rangle_{I,A}^{(3)} \otimes (a_1|0\rangle - (-1)^r a_0|1\rangle)_{Bi}]$$
$$\otimes [\prod_r |A_-\rangle \otimes \prod_{n-r-1}|A_+\rangle]_{B1,B2,...,B(i-1),B(i+1),...,Bn}.$$

Now Alice performs BSM on her two particles (information mode $I$ and entangled mode $A$) in the Bell basis and sends her BSM result to $i^{th}$ Bob through a classical 2-bit channel. From equation (7), it is clear that if $r$ is even, then Alice sends classical information 00, 01, 10, 11 for the BSM results 0, 1, 2, 3 respectively and $i^{th}$ Bob performs unitary transformation $I$, $\sigma_z$, $\sigma_x$, $\sigma_z\sigma_x$ respectively in order to generate the exact replica of the information state. On the other hand if $r$ is odd, then Alice sends classical information 01, 00, 11, 10 for the BSM results 0, 1, 2, 3 respectively and $i^{th}$ Bob performs unitary transformation $\sigma_z$, $I$, , $\sigma_z\sigma_x$, $\sigma_x$ respectively in order to generate the exact replica of the information state.

## 3 Generalization of Quantum Teleportation within a quantum network for multi-qubit information state

In this section, we present the generalization of both schemes mentioned above to the teleportation of an arbitrary $m$-qubit state of the form [17],

$$|I\rangle_{I_1,I_2,....,I_m} = [a_0|\tilde{0}\rangle + ............ + a_M|\tilde{M}\rangle]_{\{I\}}, \qquad (8)$$

within the network of $n$ receivers ($n$ Bobs), where $\{I\} \equiv I_1, I_2,...., I_m$, $M \equiv 2^m - 1$ and $|\tilde{0}\rangle$, $|\tilde{1}\rangle$,....., $|\tilde{M}\rangle$ are a binary form of the $2^m$ states $|0....^m...0\rangle, |0....^m...1\rangle,....., |1....^m...1\rangle$. Let us consider the quantum network consisting of $(n+1)$ observers, Alice and $n$ numbers of Bobs, sharing an $m(n+1)$-partite entangled state of the form,

$$|E\rangle_{\{A\},\{B1\},....,\{Bn\}} = \frac{1}{2^{m/2}}[\{\prod_{n+1}|\tilde{0}\rangle + ............. + \prod_{n+1}|\tilde{M}\rangle\}]_{\{A\},\{B1\},....,\{Bn\}}, \qquad (9)$$

where $\{A\} \equiv A_1, A_2, ...., A_m$, $\{Bj\} \equiv B_1^j, B_2^j, ......, B_m^j$ ($j$=1, 2, ...., $n$) and $\prod_{n+1}$ represents the direct product of ($n$+1) identical factors.

The initial state of the complete system is written as

$$|\psi\rangle_{\{I\},\{A\},\{B1\},....,\{Bn\}} = |I\rangle_{\{I\}} \otimes |E\rangle_{\{A\},\{B1\},....,\{Bn\}}. \tag{10}$$

Information state in modes $\{I\}$ and entangled modes $\{A\}$ are with Alice, while the entangled modes $\{Bi\}$ ($i$=1, 2, ...., $n$) are with $i^{th}$ Bob. Let Alice chooses $i^{th}$ Bob to whom information state (8) is to be teleported. For this purpose, in the first scheme, Alice asks all Bobs except the $i^{th}$ Bob to perform C-NOT operations on their each qubit with control as her qubit such that mode $A_k$ controls mode $B_k^j$ ($k$=1, 2, ...., $m$). On doing this, the equation of state (10) takes the form,

$$|\psi\rangle_{\{I\},\{A\},\{B1\},....,\{Bn\}}$$
$$= |I\rangle_{\{I\}} \otimes \frac{1}{2^{m/2}} [\{|\tilde{0}\rangle \otimes |\tilde{0}\rangle + ............ + |\tilde{M}\rangle \otimes |\tilde{M}\rangle\}]_{\{A\},\{Bi\}} \tag{11}$$
$$\otimes [\prod_{n-1} |\tilde{0}\rangle]_{\{B1\},...,\{B(i-1)\},\{B(i+1)\},...,\{Bn\}}.$$

This equation can be simplified further as

$$|\psi\rangle_{\{I\},\{A\},\{B1\},....,\{Bn\}}$$
$$= \frac{1}{2^m} [\sum_{p=0}^{2^{2m}-1} |E\rangle^{(p)}_{\{I\},\{A\}} \otimes U_p |I\rangle_{\{Bi\}}] \otimes [\prod_{n-1} |\tilde{0}\rangle]_{\{B1\},...,\{B(i-1)\},\{B(i+1)\},...,\{Bn\}}, \tag{12}$$

where states $\{|E\rangle^{(p)}_{\{I\},\{A\}}\}$ form an orthonormal basis in a $2^{2m}$ dimensional Hilbert space and $U_p^\dagger$'s are the unitary transformations. The states $\{|E\rangle^{(p)}_{\{I\},\{A\}}\}$ are called generalized Bell states [17] and are given as

$$|E\rangle^{(p)}_{\{I\},\{A\}} = \frac{1}{2^{m/2}} [(\sigma^{p_1} \otimes \sigma^{p_2} \otimes ...... \otimes \sigma^{p_m}) \tag{13}$$
$$\{|\tilde{0}\rangle \otimes |\tilde{0}\rangle + ............ + |\tilde{M}\rangle \otimes |\tilde{M}\rangle\}]_{\{I\},\{A\}},$$

where each $\sigma^{p_k}$ is unitary matrix belonging to the set $\{I, \sigma_z, \sigma_x, \sigma_x\sigma_z\}$ and acts on modes $\{A_k\}$ for $k$ = 1, 2, ...., $m$. The generalized Bell states $\{|E\rangle^{(p)}_{\{I\},\{A\}}\}$, by definition, are tensor product of $n$ standard bipartite Bell states. $U_p$ is the unitary matrix $(\sigma^{p_1} \otimes \sigma^{p_2} \otimes ....... \otimes \sigma^{p_m})$ and $p$ is decimal conversion of quaternary number $(p_1 p_2....p_m)$.

Now Alice performs generalized $2m$-qubit BSM in the basis $\{|E\rangle^{(p)}_{\{I\},\{A\}}\}$ and conveys her measurement result $p \in (0,1,...,2^{2m}-1)$ to $i^{th}$ Bob through a classical $2m$-bit channel. Then $i^{th}$ Bob performs the required unitary transformation in order to generate the exact replica of the original information state. From equation (12) it is clear that, on the basis of the measurement result '$p$', $i^{th}$ Bob performs corresponding unitary transformation $U^{\dagger}_p$ on his qubits in order to generate an exact replica of the original information state.

In the second scheme, Alice asks to all Bobs except the $i^{th}$ Bob to perform projective measurement on their each particle in the basis $|A_\pm\rangle$ (defined in section 1) and convey their results back. Let the measurement result of $r_k$ number of Bobs is $|A_-\rangle$ and rest $(n-r_k-1)$ number of Bobs is $|A_+\rangle$ on their $k^{th}$ qubit $k \in [1,2,...,m]$. On such measurement, the initial state (10) of the system becomes,

$$|\psi\rangle_{\{I\},\{A\},\{B1\},.........,\{Bn\}}$$
$$= |I\rangle_{\{I\}} \otimes \frac{1}{2^{m/2}} [\prod_2 |\tilde{0}\rangle + (-1)^{r_m} \prod_2 |\tilde{1}\rangle + (-1)^{r_{m-1}} \prod_2 |\tilde{2}\rangle + (-1)^{r_{m-1}+r_m} \prod_2 |\tilde{3}\rangle \quad (14)$$
$$+............+(-1)^{r_1+...+r_m} \prod_2 |\tilde{M}\rangle]_{\{A\},\{Bi\}}$$
$$\otimes [\prod_{r_1} |A_-\rangle \otimes \prod_{n-r_1-1} |A_+\rangle \otimes ...... \otimes \prod_{r_m} |A_-\rangle \otimes \prod_{n-r_m-1} |A_+\rangle]_{\{B1\},....,\{B(i-1)\},\{B(i+1)\},.....,\{Bn\}}.$$

This equation can be simplified further using the generalized Bell states $\{|E\rangle^{(p)}_{\{I\},\{A\}}\}$ as

$$|\psi\rangle_{\{I\},\{A\},\{B1\},.........,\{Bn\}}$$
$$= \frac{1}{2^m} [\sum_{p=0}^{2^{2m}-1} |E\rangle^{(p)}_{\{I\},\{A\}} \otimes U'_p |I\rangle_{\{Bi\}}] \quad (15)$$
$$\otimes [\prod_{r_1} |A_-\rangle \otimes \prod_{n-r_1-1} |A_+\rangle \otimes ...... \otimes \prod_{r_m} |A_-\rangle \otimes \prod_{n-r_m-1} |A_+\rangle]_{\{B1\},....,\{B(i-1)\},\{B(i+1)\},.....,\{Bn\}},$$

where $\{|E\rangle^{(p)}_{\{I\},\{A\}}\}$ are given by equation (13) and $U'_p$'s are the unitary transformations. Now Alice performs generalized $2m$-qubit BSM in the basis $\{|E\rangle^{(p)}_{\{I\},\{A\}}\}$. Depending on the result of generalized BSM and measurement results conveyed by all Bobs other than $i^{th}$ Bob, Alice conveys classical $2m$-bit information to $i^{th}$ Bob through a classical channel and then $i^{th}$ Bob performs the required unitary transformation in order to generate the exact replica of the original information state. From equation (14) and (15), it is clear that, if all $r_k$, $k \in [1,2,...,m]$, are even, then for generalized BSM result '$p$', $i^{th}$ Bob performs unitary transformation $U'^{\dagger}_p = U^{\dagger}_p$ on his qubits in order to generate an exact replica of the original information state. In other cases, when $r_k$ is odd, then Alice flips the $(2k-1)^{th}$ bit of the classical $2m$-bit information and then sends it to Bob, who performs corresponding unitary transformation on his qubits.

**4 Experimental scheme for the generation of ($n$+1)-partite GHZ state**

In order to experimentally realize the both QT schemes discussed in the Section **2**, we need to have the ability of generation of the entangled channel, *i.e.*, ($n$+1)-partite GHZ state (equation (1)). Hence our next aim is to generate the ($n$+1)-partite GHZ state. Authors [20] have shown the generation of tripartite GHZ state. Here, we extend this scheme to generate the {($n$+1)=$N$}-partite GHZ state. We consider the interaction of $\Lambda$-type three-level atom with optical coherent field. For $\Lambda$-type three level atom, level configuration of atom is shown in Figure **3**. $|0\rangle$ and $|1\rangle$ are two degenerate ground levels and $|2\rangle$ is excited level. $\omega_c$ is frequency of optical coherent field and it is largely detuned from atomic transition frequency $\omega_0$, *i.e.*, detuning $\Delta = \omega_0 - \omega_c$ is large. In large detuning limit, level $|2\rangle$ can be neglected during the interaction and effective Hamiltonian can be expressed as [21]

$$H = -\lambda a^\dagger a [|0\rangle\langle 0| + |0\rangle\langle 1| + |1\rangle\langle 0| + |1\rangle\langle 1|], \tag{16}$$

where, $\lambda = g^2/\Delta$. In reference [20], it is shown that if the interaction time is set to $\pi/2\lambda$, then we get following evolutions,

$$|0, \alpha\rangle \xrightarrow{\lambda t = \pi/2} \frac{1}{2}[|0, +\rangle - |1, -\rangle],$$

$$|1, \alpha\rangle \xrightarrow{\lambda t = \pi/2} \frac{1}{2}[|1, +\rangle - |0, -\rangle], \tag{17}$$

$$|0, +\rangle \to |0, +\rangle; \quad |1, +\rangle \to |1, +\rangle,$$

$$|0, -\rangle \to -|1, -\rangle; \quad |1, -\rangle \to -|0, -\rangle, \tag{18}$$

where $|\alpha\rangle$ and $|-\alpha\rangle$ are coherent states of radiation and $|\pm\rangle = [|\alpha\rangle \pm |-\alpha\rangle]$ are unnormalized even and odd coherent states. Let us consider a cavity $C$ in optical coherent field state $|\alpha\rangle_C$ and $N$ atoms in modes $A_1, A_2, \ldots, A_N$ initially in the states $|0\rangle_{A_1}, |0\rangle_{A_2}, \ldots, |0\rangle_{A_N}$ respectively. The initial state of the atom-cavity system is,

$$|\psi(0)\rangle_{A_1, A_2, \ldots, A_N, C} = |00\ldots 0\rangle_{A_1, A_2, \ldots, A_N} \otimes |\alpha\rangle_C. \tag{19}$$

The first atom $A_1$ is injected into the cavity $C$ to interact with the coherent field for interaction time $t = \pi/2\lambda$, then atoms $A_2, \ldots, A_N$ are sent one by one through the cavity $C$ such that each atom interact with cavity field for the time $t = \pi/2\lambda$, the state of the system evolves according to the evolutions in equation (17-18) and we get,

$$|\psi\{N\pi/2\lambda\}\rangle_{A_1,A_2,...,A_N,C}$$

$$= \frac{1}{2}[|00....0,+\rangle + (-1)^N |11....1,-\rangle]_{A_1,A_2,...,A_N,C},$$

$$= \frac{1}{2}[\{|00....0\rangle + (-1)^N |11....1\rangle\} \otimes |\alpha\rangle$$

$$+ \{|00....0\rangle - (-1)^N |11....1\rangle\} \otimes |-\alpha\rangle]_{A_1,A_2,...,A_N,C}.$$

Now we perform displacement $D(\alpha)$ into the cavity $C$. This results the states,

$$|\alpha\rangle \xrightarrow{D(\alpha)} |2\alpha\rangle, \quad |-\alpha\rangle \xrightarrow{D(\alpha)} |v\rangle,$$

where $v$ corresponds to vacuum and we get the final state,

$$|\psi\{N\pi/2\lambda\}\rangle_{A_1,A_2,...,A_N,C}$$

$$= \frac{1}{2}[\{|00....0\rangle + (-1)^N |11....1\rangle\} \otimes |2\alpha\rangle \quad (20)$$

$$+ \{|00....0\rangle - (-1)^N |11....1\rangle\} \otimes |v\rangle]_{A_1,A_2,...,A_N,C}.$$

If we perform photon counting measurement (PCM) into the cavity $C$, we get two results, either non-zero photon is detected or zero photon is detected. The discrimination between non-zero photon and zero photon can be done successfully for appreciable mean number of photons $|\alpha|^2$. In case of non-zero PCM result, the generated state is

$$\frac{1}{\sqrt{2}}[|00....0\rangle + (-1)^N |11....1\rangle\}]_{A_1,A_2,...,A_N}, \quad (21)$$

and when result of PCM into the cavity $C$ is zero, the generated state is

$$\frac{1}{\sqrt{2}}[|00....0\rangle - (-1)^N |11....1\rangle\}]_{A_1,A_2,...,A_N}. \quad (22)$$

Depending on $N$ is even or odd, both the above states, given by equations (21) and (22), can be converted into the required $N$-partite GHZ state (1) by local operations. Hence, for both PCM results, the required $N$-partite GHZ state (1) is generated.

Now we consider the experimental feasibility of this scheme. We consider the cavity [22] made up of two carefully polished niobium mirrors of diameter D=50 mm facing each other and the field mode occupies only 10% the size of cavity (i. e. 5mm). For this cavity, the required atom-cavity-field interaction time is $t \approx 10^{-4}$ s and then time required by atom to cross the cavity is $T \approx 10^{-3}$ s and the cavity damping time is $T_D \approx 1$ s [20]. If we consider that oven (source of atoms) and cavity $C$ are 10mm apart to each other in vacuum space, then total flight time of $N$ atoms is $(7T/5) + (N-1)t$. Since cavity damping time is $T_D \approx 1$ s, we can take total flight time of atoms $\approx 10^{-1}$ s. For this time, the maximum value of $N$ comes out to nearly $10^3$. Thus, in this scheme, we have a restriction on the number of atoms and hence, we can have a quantum network consisting of utmost $10^3$ parties.

To remove this restriction, we now present another experimental scheme. Since this scheme requires the number of cavities equal to the number of the atoms in the required GHZ state, this scheme consumes more resources than the earlier proposed scheme in this section. Let us prepare $N$ atoms in modes $A_1$, $A_2$, ....., $A_N$ initially in the states $|0\rangle_{A_1}$, $|0\rangle_{A_2}$, ....., $|0\rangle_{A_N}$ respectively and $N$ cavities $C_1$, $C_2$, ....., $C_N$ in optical coherent field state $|\alpha\rangle_{C_1}$, $|\alpha\rangle_{C_2}$, ....., $|\alpha\rangle_{C_N}$ respectively. The initial state of the atom-cavity system is,

$$|\psi(0)\rangle_{A_1,A_2,A_3,.....,A_N,C_1,C_2,C_3,.....,C_N} \\ = |000.....0\rangle_{A_1,A_2,A_3,.....,A_N} |\alpha\alpha\alpha.....\alpha\rangle_{C_1,C_2,C_3,.....,C_N}. \qquad (23)$$

Our detailed scheme is shown in figure **3**. The atoms $A_1$, $A_2$, ...., $A_N$ are sent simultaneously through the cavities $C_1$, $C_2$,...., $C_N$ respectively to interact with the cavity field. For each atom, if the interaction time satisfies $t = \pi/2\lambda$, then the state of the atom-cavity system evolves according to the evolution in equation (17) and we get,

$$|\psi(\pi/2\lambda)\rangle_{A_1,A_2,A_3,.....,A_N,C_1,C_2,C_3,.....,C_N} \\ = \frac{1}{2^N}(|0,+\rangle - |1,-\rangle)_{A_1,C_1}(|0,+\rangle - |1,-\rangle)_{A_2,C_2} \qquad (24) \\ (|0,+\rangle - |1,-\rangle)_{A_3,C_3}.............(|0,+\rangle - |1,-\rangle)_{A_N,C_N}.$$

After this, the atoms $A_1$, $A_2$,...., $A_N$ are sent simultaneously through the cavities $C_2$, $C_3$,...., $C_1$ respectively. If still the interaction time satisfies $t = \pi/2\lambda$ for each atom, then the state of the atom-cavity system evolves according to the evolution in equation (18) and we get,

$$|\psi(\pi/\lambda)\rangle_{A_1,A_2,A_3,.....,A_N,C_1,C_2,C_3,.....,C_N} \\ = \frac{1}{2^N}[|000....0\rangle(|+++....+\rangle + |---....-\rangle) \\ + |000....1\rangle(|+++....-\rangle + |---....+\rangle) \qquad (25) \\ + |000....10\rangle(|+++....-+\rangle + |---....+-\rangle) \\ + |000....11\rangle(|+++....--\rangle + |---....++\rangle) \\ + ........]_{A_1,A_2,A_3,.....,A_N,C_1,C_2,C_3,.......,C_N}.$$

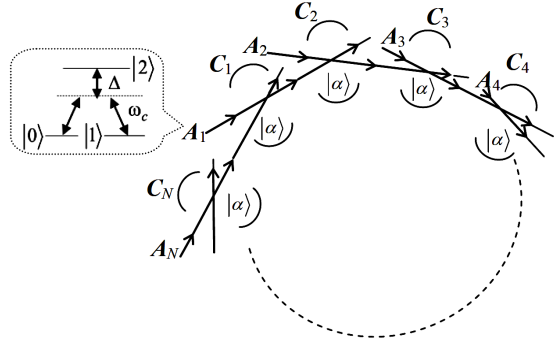

**Figure 3**: Scheme for generation of *N*-partite GHZ state (equation (6.1). Callout represents the level configuration of $\Lambda$-type three level atoms. $|0\rangle$ and $|1\rangle$ are two degenerate levels and $|2\rangle$ is excited level. All cavities $C_1$, $C_2$, ....., $C_N$ are initially prepared in optical coherent field state $|\alpha\rangle$, while all atoms $A_1$, $A_2$, ...., $A_N$ are initially in the state $|0\rangle$.

Now all atoms $A_1$, $A_2$, ..... and $A_N$ are measured in the computational basis $\{|0\rangle, |1\rangle\}$. For each measurement result, we get a GHZ state of radiation involving $N$ modes. If after measurement all atoms are in the state $|0\rangle$, then from equation (25), the generated GHZ state of radiation is,

$$|\psi(\pi/\lambda)\rangle_{C_1,C_2,C_3,....,C_N} = \frac{1}{2^{N/2}}[|+++....+\rangle + |---....-\rangle]_{C_1,C_2,C_3,....,C_N}. \qquad (26)$$

For other measurement results, one can find the generated GHZ state of radiation. Let us suppose that after measurement, we get the generated GHZ state of radiation given by equation (26). Now we take $N$ atoms $A_1', A_2', A_3',...., A_N'$ initially in the state $|0\rangle_{A_1'}, |0\rangle_{A_2'}, ....., |0\rangle_{A_N'}$ and the state of system is

$$|\psi(0)\rangle_{A_1',A_2',A_3',....,A_N',C_1,C_2,C_3,....,C_N}$$
$$= \frac{1}{2^{N/2}}|000....0\rangle_{A_1',A_2',A_3',...,A_N'}[|+++....+\rangle + |---....-\rangle]_{C_1,C_2,C_3,....,C_N}. \qquad (27)$$

The atoms $A_1', A_2', A_3',...., A_N'$ are injected simultaneously into the cavities $C_1$, $C_2$,...., $C_N$ respectively. If we control the velocities of atoms such that the interaction time satisfies $t = \pi/2\lambda$ for each atom, then the state (26) of the atom-cavity system evolves according to the evolution in equation (18) and we get,

$$|\psi(\pi/2\lambda)\rangle_{A'_1,A'_2,A'_3,\ldots,A'_N,C_1,C_2,C_3,\ldots,C_N}$$

$$= \frac{1}{2^{N/2}}[|000\ldots0\rangle|+++\ldots+\rangle \tag{28}$$

$$+(-1)^N|111\ldots1\rangle|---\ldots-\rangle]_{A'_1,A'_2,A'_3,\ldots,A'_N,C_1,C_2,C_3,\ldots,C_N}.$$

In terms of coherent states $|\alpha\rangle$ and $|-\alpha\rangle$, the state (28) can be rewritten as

$$|\psi(\pi/2\lambda)\rangle_{A'_1,A'_2,A'_3,\ldots,A'_N,C_1,C_2,C_3,\ldots,C_N}$$

$$= \frac{1}{2^{N/2}}[(|000\ldots0\rangle+(-1)^N|111\ldots1\rangle)(|\alpha,\alpha,\alpha\ldots\alpha\rangle+|\alpha,\alpha,\alpha\ldots-\alpha,-\alpha\rangle+\ldots\ldots) \tag{29}$$

$$+(|000\ldots0\rangle-(-1)^N|111\ldots1\rangle)(|\alpha,\alpha,\alpha\ldots-\alpha\rangle+|\alpha,\alpha,\alpha\ldots-\alpha,\alpha\rangle+\ldots\ldots$$

$$\ldots)]_{A'_1,A'_2,A'_3,\ldots,A'_N,C_1,C_2,C_3,\ldots,C_N}.$$

In the first and second line of the right hand side of the above equation, number of states $|-\alpha\rangle$ are even and odd respectively.

From equation (29), it is clear that the terms containing even number of $|-\alpha\rangle$ are with the *N*-partite GHZ state,

$$\frac{1}{\sqrt{2}}(|000\ldots0\rangle+(-1)^N|111\ldots1\rangle)_{A'_1,A'_2,A'_3,\ldots,A'_N}, \tag{30}$$

while the terms containing odd number of $|-\alpha\rangle$ are with the *N*-partite GHZ state,

$$\frac{1}{\sqrt{2}}(|000\ldots0\rangle-(-1)^N|111\ldots1\rangle)_{A'_1,A'_2,A'_3,\ldots,A'_N}. \tag{31}$$

Also since these two *N*-partite GHZ states are convertible into each other using local operations, hence it is enough to generate any of the two. For this purpose, we now make use of displacement operator $D(\alpha)$ into each cavity $C_1, C_2,\ldots, C_N$. Equation (29) thus becomes,

$$|\psi(\pi/2\lambda)\rangle_{A'_1,A'_2,A'_3,\ldots,A'_N,C_1,C_2,C_3,\ldots,C_N}$$

$$= \frac{1}{2^{N/2}}[(|000\ldots0\rangle+(-1)^N|111\ldots1\rangle)(|2\alpha,2\alpha,2\alpha\ldots2\alpha\rangle+|2\alpha,2\alpha,2\alpha\ldots v,v\rangle+\ldots\ldots) \tag{32}$$

$$+(|000\ldots0\rangle-(-1)^N|111\ldots1\rangle)(|2\alpha,2\alpha,2\alpha\ldots v\rangle+|2\alpha,2\alpha,2\alpha\ldots v,2\alpha\rangle+\ldots\ldots$$

$$\ldots)]_{A'_1,A'_2,A'_3,\ldots,A'_N,C_1,C_2,C_3,\ldots,C_N}.$$

Now PCM is done in each cavity. Also for appreciable mean number of photons $|\alpha|^2$ in coherent state, we can discriminate between non-zero photon and zero photon successfully. If the result of PCM in even number of cavities is zero, the generated *N*-partite GHZ state is given by equation (30). When result of PCM in odd number of cavities is zero, the generated *N*-partite GHZ state is given by equation (31). Further, depending on the value of *N* (even or odd), one can get the required GHZ state (1) because the two states, obtained after PCM, are convertible into each other by local operations.

Since this scheme requires only three interactions and a measurement on atoms and radiation modes. Three interactions take total time $\approx 3T = 3 \times 10^{-3}$ s, which is much less than the cavity damping time $T_D \approx 1$ s.

## 5. Conclusions and Discussion

We considered the problem of teleporting an unknown information state within the quantum network by a sender, say, Alice to any given receiver out of several receivers, say, Bob1, Bob2 etc. We suggested two schemes for QT of single qubit information state within the quantum network. The first scheme involved C-NOT operation on Bob's qubit with control as Alice's qubit and the second scheme involved a measurement in the basis $\frac{1}{\sqrt{2}}[|0\rangle \pm |1\rangle]$ by Bobs on their qubits. We generalized both schemes for quantum teleportation of multi-qubit information state within the quantum network.

Further, we suggested two efficient schemes for the generation of (*n*+1)-partite GHZ state, which is used as quantum channel in both schemes, using interaction between highly detuned Λ-type three-level atoms and optical coherent field. We also discussed the possible experimental feasibility our schemes and found that these scheme can be realized with presently available technology.

In this paper, we considered the quantum network composed of a sender and several receivers. This study can be generalized further for some other type of quantum networks. For example, quantum network may be composed of several senders and one receiver, in which out of several senders, one sends the quantum information to the receiver. This can be achieved simply by replacing the all Bobs (receivers) into Alices (senders) and Alice (sender) into Bob (receiver). In this quantum network, suppose Bob needs the quantum information state of *i*<sup>th</sup> Alice out of all Alices. For this, *i*<sup>th</sup> Alice teleports her quantum information state to Bob.

There may be another quantum network, which consists of several sender and several receivers, *i.e.*, *m* senders and *n* receivers. Out of *m* senders, one sends the quantum information state to a selected receiver who requires the quantum information state.

## Acknowledgements

We are thankful to *Prof. N. Chandra* and *Prof. R. Prakash* for their interest in this work. We would like to thank *Dr. D. K. Singh*, *Dr. D. K. Mishra*, *Dr. R. Kumar*, *Dr. P. Kumar*, *Mr. Ajay K. Yadav* and *Mr. Vikram Verma* for helpful and stimulating discussions.

Author AKM acknowledges financial support of CSIR under CSIR-SRF scheme and author MKM acknowledges financial support of UGC under UGC-SRF fellowship scheme.